\documentclass[10pt,twocolumn,twoside]{IEEEtran}

\usepackage{xcolor}
\usepackage{graphicx}
\usepackage{tikz}
\usepackage{color}
\usepackage{amsmath}
\usepackage{subfig}
\usepackage{algorithm}
\usepackage{algorithmic}
\usepackage{amssymb}
\graphicspath{{./Fig/}}
\usepackage{math}
\usepackage{caption}
\usepackage{array}
\usepackage{enumitem}
\usepackage{balance}
\usepackage{url}

\usepackage[final]{review}
\setcoverletter{coverletter-R1.tex}
\setrevision{1}

\begin{document}


\title{Narrow-band Deep Filtering for \\ Multichannel Speech Enhancement 
}
\author{Xiaofei Li and Radu Horaud
\thanks{X. Li is with Westlake University, China, and R. Horaud is with Inria Grenoble Rh\^one-Alpes and with Univ. Grenoble Alpes, France. }
%\thanks{This work was supported by the ERC Advanced Grant VHIA \#340113.}
}

%\markboth{IEEE/ACM Transactions on Audio, Speech and Language Processing}{X. Li and R. Horaud: Narrow-band Deep Filtering for Multichannel Speech Enhancement}

\maketitle
\begin{abstract}
In this paper we address the problem of multichannel speech enhancement in the short-time Fourier transform (STFT) domain. 
A long short-time memory (LSTM) network takes as input a  sequence of STFT coefficients associated with a frequency bin of multichannel noisy-speech signals. The network's output is the corresponding sequence of single-channel cleaned speech.  We propose several clean-speech network targets, namely, the magnitude ratio mask, the complex STFT coefficients and the (smoothed) spatial filter. A prominent feature of the proposed model is that the same LSTM  architecture, with identical parameters, is trained across frequency bins. The proposed method is referred to as narrow-band deep filtering. This choice stays in contrast with traditional wide-band speech enhancement methods. The proposed deep filtering is able to discriminate between speech and noise by exploiting their different temporal and spatial characteristics: speech is non-stationary and spatially coherent while noise is relatively stationary and weakly correlated across channels. 
This is similar in spirit with unsupervised techniques, such as spectral subtraction and beamforming. We describe extensive experiments with both mixed signals (noise is added to clean speech) and real signals (live recordings). We empirically evaluate the proposed architecture variants using speech enhancement and speech recognition metrics, and we compare our results with the results obtained with several state of the art methods. In the light of these experiments we conclude that narrow-band deep filtering has very good speech enhancement and speech recognition performance, and excellent generalization capabilities in terms of speaker variability and noise type. 

\end{abstract}


\section{Introduction}
This paper addresses the problem of multichannel speech enhancement/denoising using deep learning. 
In recent years, speech enhancement based on deep neural networks has been thoroughly and successfully investigated, see \cite{wang2018} for an overview. These methods are often conducted in the time-frequency (TF) domain, and can be broadly categorized into either monaural or multichannel techniques. The monaural techniques use a neural network to map noisy-speech spectral features onto clean speech targets. The input features, e.g. (logarithm) signal spectra, cepstral coefficients, or linear prediction based features, generally represent the frame-wise full-band spectral structure associated with noisy speech. The target consists of either clean speech spectral features or of ideal binary/ratio masks (IBM/IRM) which are subsequently applied to the noisy-speech input. Recovering the clean phase is beneficial for improving the perceptual quality, which however is difficult due to the lack of a clear structure for phase spectra. Alternatively, the real and image part of the speech spectra both show a clear spectral structure, which are thus used either to construct a complex IRM (cIRM) \cite{williamson2016} or  directly as the target \cite{fu2017,wang2020}. Widely used neural architectures for speech enhancement include feed-forward neural networks (FNNs), convolutional neural networks (CNNs) and recurrent neural networks (RNNs). The temporal dynamics of speech can be modeled by stacking context frames in the FNN input, or by dilated CNN \cite{tan2019}, while they are automatically modeled by RNNs. In \cite{weninger2014,chen2017}, the memory-enhanced RNN, i.e. LSTM, is used to learn the long-term dependencies of signals. There are only a handful of methods that process frequency bands separately, e.g. \cite{wang2013,jiang2014}, namely a neural network is trained for each subband: these subband spectral features are mapped onto subband targets.

As for multichannel speech enhancement, it is popular to combine supervised monaural approaches with unsupervised beamforming methods, e.g. \cite{heymann2016,boeddeker2018}. The output of the former, i.e. a TF mask, is used to discriminate between speech and noisy TF units, based on which the steering vector of desired speech and noise covariance matrices are computed by the latter. These approaches don't learn the spatial information. To exploit the spatial information, interchannel features (sometimes combined with spectral features), e.g. interaural time-, phase-, and level-difference (ITD, IPD and ILD) and the cross-correlation function (CCF), are input to a neural network either for full-band TF mask prediction, e.g. \cite{zhang2017deep,wang2019,yoshioka2019}, or for subband TF mask prediction, e.g. \cite{jiang2014,pertila2015}. Due to the use of the interchannel features, these methods are sensitive to the position of the speech source. Therefore, they consider the position of the speech source to be fixed or to be known. In \cite{chakrabarty2019}, the magnitude and phase of the short-time Fourier transform (STFT) coefficients of all frequency bands and microphones are directly input to a convolutional recurrent neural network (CRNN), and predict the monaural full-band TF masks, where the convolutional layers extract the inter-channel information and the recurrent layers learn the temporal dynamics. This method is designed to discriminate between the spatial characteristics of directional speech sources and diffuse or uncorrelated sources, i.e. noise, and it is not sensitive to the position of the speech source. In the above multichannel techniques, TF masks serve as a preliminary of a beamformer-based estimator. Even though TF masking is able to improve the speech perceptual quality, it is widely accepted that the signal artifacts created by masking, more specifically by the nonlinear operation of masking, is harmful for automatic speech recognition (ASR). Therefore, beamforming is generally used as an interface between the speech enhancement/separation front-end and the ASR back-end.
There are several attempts to skip the masking step, and directly predict the beamformer using network. In \cite{xiao2016}, an FNN is designed to learn the frequency-domain beamformers from the time-domain generalized cross correlation (GCC) function. In \cite{li2016neural}, the time-domain spatial filters are adaptively predicted by inputting the network a segment of multichannel raw waveforms.  The beamforming network proposed in \cite{meng2017} takes as input the full-band multichannel STFT coefficients and predicts the beamformers for all the frequency bands.  

In this work, we propose a LSTM-based multichannel speech denoising method. Unlike the vast majority of existing approaches that perform full-band speech enhancement, the proposed method processes each STFT frequency bin separately: this is referred to as narrow-band (or frequency-wise) deep filtering. The proposed LSTM training is performed with input and target sequences of noisy- and clean-speech, respectively. Each input is a sequence of multichannel STFT coefficients associated with a single frequency bin. Correspondingly, the target is a sequence of clean speech taken at the same frequency for the reference channel. Importantly, the network weights are shared across frequency bins, which encourages the network to learn common information accross frequency bins, and also leads to a dramatic reduction in the complexity and computational burden of the training process.  Our approach is grounded by the fact that a large number of unsupervised speech enhancement methods exploit frequency-wise narrow-band information. More precisely, the proposed method is motivated on the following grounds:
\begin{itemize}
\item
The frequency-wise temporal evolution of the STFT magnitude is informative due to the non-stationary nature of speech against the stationarity of noise, which stands at the foundation of unsupervised single-channel noise power estimation, e.g. \cite{li2016icassp,gerkmann2012}, as well as multichannel relative transfer function (RTF) estimation \cite{gannot2001,li2015icassp}. Recently it was demonstrated that a LSTM network is able to accomplish monaural frequency-wise noise power estimation \cite{li2019spl};

\item
The frequency-wise spatial characteristics of the STFT coefficients fully reflect both the directionality of speech and the diffusion of noise and reverberation. This is the foundation of speech enhancement methods such as the coherent-to-diffuse power ratio method \cite{schwarz2015} and beamforming techniques \cite{gannot2001,gannot2017}. Moreover, the temporal dynamics of frequency-wise spatial correlation contain motion information associated with a speech source;

\item
The frequency-wise representation is informative for clean-speech estimation. Indeed, with proper parameter estimation, single-channel spectral subtraction (Bayesian filtering) \cite{ephraim1984,cohen2001}, and multichannel spatial filtering, e.g. beamforming \cite{gannot2001} and multichannel Wiener filtering \cite{brandstein2013}, are performed independently across frequencies. 
\end{itemize}
Overall, the proposed LSTM architecture is expected to fully exploit the frequency-wise information, not only by learning a regression from the input sequence to the output sequence, but also by learning a group of functions for clean speech estimation. By sharing the network weights  across frequencies, the network is encouraged not to learn the subband spectral structure of siganls, but to learn the narrow-band information mentioned above, and to perform narrow-band deep filtering. The proposed method is similar to \cite{chakrabarty2019} in that the network learns how to discriminate between the spatial characteristics of directional speech sources and the diffuse/uncorrelated nature of noise, hence the method is agnostic to the position of the speech source.

%Compared to other subband techniques, that learn a different network for each frequency band, e.g. \cite{wang2013,jiang2014}, the proposed method learns a common network for all frequency bands, which encourages to learn information that is shared across the frequency. This is similar in spirit with unsupervised methods.
Compared to full-band techniques \cite{williamson2016,fu2017,wang2020,tan2019,weninger2014,chen2017,heymann2016,boeddeker2018,zhang2017deep,
wang2019,yoshioka2019,pertila2015,chakrabarty2019,meng2017}, the proposed method ignores cross-band information, and focuses on learning narrow-band information. On one hand, this indeed loses some useful informations, such as the spectral information. On the other hand, it has the following advantages: (i)~it is questionable whether full-band  models are able to learn the narrow-band information mentioned above. As shown below, by focusing on the narrow-band signal representations, the proposed method is able to learn long-term temporal dependencies, e.g. on the order of 150 STFT frames; (ii)~due to the reduced dimension of both the input and the output, the proposed network has a smaller number of parameters than full-band models, and hence it requires much less training data and both training and prediction have a lower computational cost; (iii)~the proposed method is not sensitive to the wide-band spectral pattern of signals, since it only exploits the narrow-band information. As a result, the proposed network has a very good generalization capability in terms of speaker variability and noise type, and  (iv) experiments demonstrate that the enhanced speech obtained with the proposed method can be directly used for ASR, which means the signal artifacts caused by the prediction error of the proposed narrow-band network are not detrimental for ASR. 

This paper is an extended version of a recently published conference paper \cite{li2019waspaa}, in which we proposed a narrow-band LSTM architecture for speech enhancement and we demonstrated  its effectiveness when using the magnitude ratio mask as a network target. \addnote[contributions]{1}{The contributions of this work over \cite{li2019waspaa} include: 
\begin{itemize}
\item In addition to the magnitude ratio mask, we evaluate other targets, namely the STFT complex coefficient and a spatial filter. These two targets are not new on their own, as the complex STFT coefficient has been used in \cite{fu2017,wang2020}, and beamformer in \cite{xiao2016,li2016neural,meng2017}. However, the theoretical bases for estimating them in this work are completely different from the ones in other works: (i) the prediction of the complex STFT coefficient in \cite{fu2017,wang2020} is based on the fact that the real and image spectrograms both have a clear structure being similar to the magnitude spectrogram, and thus can be predicted based on supervised regression. In contrast, in the proposed narrow-band method, the spectrogram structure obviously does not exist. Instead, we aim to exploit the spatial features of signals to estimate the complex STFT coefficient of clean speech; (ii) in the previous deep beamforming techniques \cite{xiao2016,li2016neural,meng2017}, the beamformer of all the frequencies are predicted together by one single network. However, such setup has never been testified, as the unsupervised beamformer \cite{gannot2001,gannot2017} is usually estimated frequency-wise. The beamforming techniques consists of two components, i.e. parameter (such as RTF and noise covariance matrix) estimation and beamformer computation. Narrow-band has rich information for parameter estimation as discussed above, and beamformer computation is naturally conducted frequency-wise. Therefore, the proposed narrow-band spatial filtering technique appears to be a supervised deep-learning implementation of unsupervised beamforming techniques.  
\item The proposed method is extensively evaluated with more experiments in terms of the speaker/noise-generalization capability and speech enhancement. In addition, we evaluate the automatic speech recognition (ASR) performance of the proposed method. We do have an important new finding, namely the proposed narrow-band framework is more suitable for ASR compared to the full-band techniques, although the latter may achieve better speech enhancement evaluation scores. Different speech enhancement methods would bring different types of processing artifacts \cite{wang2019bridging}. Our experiments demonstrate that the wide-band artifacts or cross-band structured artifacts, brought by the full-band methods are more harmful than the narrow-band artifacts brought by the proposed method.
\item As for the spatial filter target, to incorporate one important characteristic of beamforming, i.e. beamformer is somehow temporally smoothed, we propose a new training loss to impose the temporal smoothing on the predicted spatial filter. This keeps the temporal consistence of both the enhanced speech and the residual noise, and thus improves the ASR performance, although the speech enhancement performance degrades. 
\end{itemize}
Overall, this work comprehensively presents and evaluates the narrow-band deep filtering method. Different targets are evaluated, which is important and necessary since the theoretical bases for estimating each target in narrow-band are different from the ones in wide-band. It is also important to evaluate the ASR performance of the proposed method, since state-of-the-art speech enhancement networks do not necessarily improve ASR performance.}  

The remainder of this paper is organized as follows. Section \ref{sec:method} describes the proposed narrow-band deep filtering model and the adopted LSTM architecture. Section \ref{sec:experiment} describes the experimental setup, the LSTM network training characteristics, the speech enhancement and speech recognition experimental results. Section \ref{sec4} concludes the paper. Supplemental material (examples of processed noisy speech utterances) are available at \url{https://team.inria.fr/perception/research/mse-lstm/}.

\section{Narrow-band Speech Enhancement Networks}
\label{sec:method}

Let the multichannel signals be represented in the STFT domain:
\begin{equation}
x_i(k,t)=s_i(k,t)+u_i(k,t), 
\end{equation}
where $x_i(k,t)$, $s_i(k,t)$ and $u_i(k,t)$ are the complex-valued STFT coefficients of the microphone, speech and noise signals, respectively, and where
$i\in\{1 \dots I\}$, $k\in\{0 \dots K-1\}$ and $t\in\{1 \dots T\}$ denote the channel (microphone),  frequency-bin and frame indices, respectively. In this paper the focus is on signal denoising task and hence the reverberation effect is not addressed. Therefore, the objective is to recover the (possibly reverberant) speech signal of one reference channel, e.g. $s_r(k,t)$, where $r$ denotes the reference channel. In the proposed method and as already mentioned, a single network is trained using the narrow-band sequences over all frequency bins, and the trained network is then used to predict a sequence at each frequency bin. Thence, for the sake of clarity, the frequency-bin index $k$ will be omitted hereafter. 

\subsection{Input Features}
For each TF bin, the real and imaginary parts, $\mathcal{R}(\cdot)$, $\mathcal{I}(\cdot)$ of the multichannel STFT coefficients are concatenated into the vector:
\begin{align}\label{eq:inputfeatures}
\mathbf{x}(t) =  \big(\mathcal{R}(x_1(t)),\mathcal{I}(x_1(t)),\dots, \mathcal{R}(x_I(t)),\mathcal{I}(x_I(t))\big)^{\top},
\end{align}
where $^\top$ denotes vector transpose.  $\mathbf{x}(t)\in\mathbb{R}^{2I} $ contains information associated with one TF bin. 
The input sequence of LSTM is a temporal sequence of such vectors at each frequency bin, namely:
\begin{equation}\label{eq:xseq}
\tilde{\mathbf{X}}=  \big(\mathbf{x}(1),\dots,\mathbf{x}(t),\dots,\mathbf{x}(T) \big),
\end{equation} 
where $T$ denotes the number of time steps of the LSTM network.  To facilitate network training, the input sequence has to be normalized to equalize the input levels across channels and across time. We empirically set to $1$ the STFT magnitude of the reference channel, namely:
% i.e. $\mu = \frac{1}{T}\sum_{t=1}^T |x_r(t)|$, where $|\cdot|$ denotes the module. Accordingly, the input sequence is normalized as:  
\begin{equation}\label{eq:xseqnor}
\begin{cases}
\mathbf{X} = \tilde{\mathbf{X}}/\mu \\
\mathrm{with: } \quad \mu = \frac{1}{T}\sum_{t=1}^T |x_r(t)|.
\end{cases}
\end{equation}

\subsection{Output Target and Training Loss} 
As already mentioned, we want to recover the clean speech signal of the reference channel, e.g. $s_r(t)$. To this end, we test the following network targets.
\subsubsection{Magnitude Ratio Mask (MRM)}
For each TF bin, the rectified STFT magnitude ratio mask
\begin{align}\label{eq:mrm}
M(t)=\text{min}\left(\frac{|s_r(t)|}{|x_r(t)|},1\right)
\end{align}
is the target, where the function $\text{min}(\cdot)$ rectifies the mask to fall in the range $[0,1]$. For each frequency bin, the target sequence is 
\begin{equation}\label{eq:mseq}
\mathbf{M}= \big(M(1),\dots,M(t)\dots,M(T) \big).
\end{equation} 
The mean squared error (MSE) of MRM, i.e. $({M}(t)-\hat{M}(t))^2$, is taken as the training loss, where $\hat{M}(t)$ denotes the MRM network prediction. At test, the MRM prediction $\hat{M}(t)$ is used to estimate the module of the STFT coefficient while its phase is the phase of the reference channel:
\begin{align}
| \hat{s} (t) | &=\hat{M}(t) |x_r (t)|,  \\
\arg (  \hat{s} (t) ) &= \arg ( x_r (t ) )
\end{align} 
It was demonstrated in \cite{wang2014} that, in the framework of monaural full-band masking, the MRM achieves the best performance among various magnitude-based masks, such as IBM or IRM. Our preliminary experiments within the present framework also demonstrate that this target performs slightly better than IRM. The magnitude mask performs as a spectral subtraction gain for denoising in \cite{ephraim1984,cohen2001} and for dereverberation in \cite{schwarz2015}. Many narrow-band informations can be used to estimate the gain, such as the stationarity and diffuseness of signals.  

\subsubsection{STFT Complex Coefficient (CC)}
\label{sec:cc} 
In the monaural full-band speech enhancement techniques, cIRM \cite{williamson2016} and real/image spectra \cite{fu2017,wang2020} are taken as the training targets to estimate the complex spectra, since both real and image spectrograms have a clear structure and thus can be predicted by supervised regression. In this work, the narrow-band network does not exploit the spectral structure of the signal. Instead, we estimate the speech STFT coefficient from the multi-microphone signals, which is possible: the speech images in the multi-microphone signals are actually the source speech multiplied by the acoustic transfer functions, alternatively the speech image at the reference channel multiplied by the RTFs. The RTFs are time-invariant (resp. slowly time-varying) for the static (resp. moving) speaker case, and can be (adaptively) estimated. Then, the speech STFT coefficient of the reference channel can be estimated by such as beamforming. Note that this RTF-based beamforming technique just serves here as an example to show that, at one frequency bin, it is possible to estimate the speech STFT coefficient from the multi-microphone signals. We let the network automatically learn a function to do this, by exploiting the spatial features of speech and noise.  We have compared CC and cIRM with some preliminary experiments, while similar performance were achieved. 

Formally, the real and imaginary parts of $s_r(t)$ for one TF bin, i.e. 
\begin{align}
\mathbf{s}(t)=(\mathcal{R}(s_r(t)),\mathcal{I}(s_r(t))) \in \mathbb{R}^2
\end{align} 
are directly used as the network target. For each frequency bin, the target sequence is 
\begin{equation}\label{eq:sseq}
\mathbf{S}=  \big(\mathbf{s}(1),\dots,\mathbf{s}(t)\dots,\mathbf{s}(T) \big).
\end{equation} 
According to the input sequence normalization, i.e. \eqref{eq:xseqnor}, the target signal is also normalized with $\mu$, like $s_r(k,t)/\mu$. However, we keep to use $s_r(k,t)$ to denote the normalized signal for notational simplicity. The training loss is the MSE between the STFT coefficient of clean speech and the network prediction, i.e. $\|\mathbf{s}(t)-\hat{\mathbf{s}}(t)\|^2$. 

At test, $\hat{\mathbf{s}}(t)$ is the predicted enhanced signal. The signal normalization is conducted for each frequency independently, thence the enhanced signal $\hat{\mathbf{s}}(t)$ should be multiplied by $\mu$ to keep the level consistency across frequencies. 

\subsubsection{Spatial Filtering}
\addnote[spatial1]{1}{
The combination of TF masking and beamforming techniques often achieve state-of-the-art ASR performance. The beamforming techniques consists of two components, i.e. parameter estimation and beamformer computation. For each frequency, parameters, e.g. speech RTF and noise covariance matrix, are estimated using the speech-dominant and noise-dominant TF bins, respectively, and then beamformer is derived based on some criteria. In the techniques of combining TF masking and beamforming, the monaural full-band TF masking provides an accurate classification of speech-dominant and noise-dominant TF bins, which make a great contribution to the success of these techniques. This success motivates the development of deep beamforming techniques \cite{xiao2016,li2016neural,meng2017}, which leverage one single network to directly predict the beamformer for all the frequencies. Considering that beamforming is actually a narrow-band method, namely its parameter estimation and beamformer computation are both conducted frequency-wise, it seems not reasonable to predict the beamformer for all the frequencies together. In contrast, we think the present narrow-band framework is naturally consistent to beamforming. As discussed above, narrow-band also provides rich pieces of information for speech/noise TF bins classification, such as the stationarity and spatial characteristics of signals. To explicitly mimic the beamforming-like techniques, we let the network output a multichannel spatial filter. 
}

Formally, for each TF bin, define the multichannel spatial filter $\mathbf{w}(t) \in \mathbb{R}^{2I}$ by:
\begin{align}\label{eq:spatialfilteroutput}
\mathbf{w}(t) =  \big(\mathcal{R}(w_1(t)),\mathcal{I}(w_1(t)),\dots, \mathcal{R}(w_I(t)),\mathcal{I}(w_I(t))\big)^{\top}.
\end{align}
The output is then used to estimate the clean speech,
\begin{equation}
\label{eq:spatialfilterprediction}
 \hat{\mathbf{s}}_{\text{sf}}(t)=(\mathcal{R}(\hat{s}_{\text{sf}}(t)),\mathcal{I}(\hat{s}_{\text{sf}}(t)))^T, 
 \end{equation}
 by applying the following complex-valued spatial filtering to the input:
\begin{align}
\mathcal{R}(\hat{s}_{\text{sf}}(t)) &=  \sum_{i=1}^{I} \big( \mathcal{R}(w_i(t))\mathcal{R}(x_i(t))-\mathcal{I}(w_i(t))\mathcal{I}(x_i(t)) \big), \nonumber \\
\mathcal{I}(\hat{s}_{\text{sf}}(t)) &=  \sum_{i=1}^{I} \big( \mathcal{R}(w_i(t))\mathcal{I}(x_i(t))+\mathcal{I}(w_i(t))\mathcal{R}(x_i(t)) \big). \nonumber
\end{align}
For each frequency bin, the sequence of spatial filter is 
\begin{equation}\label{eq:wseq}
\mathbf{W}=  \big(\mathbf{w}(1),\dots,\mathbf{w}(t)\dots,\mathbf{w}(T) \big).
\end{equation} 

\addnote[spatial2]{1}{
The major goal of deep spatial filtering is to make the enhanced signal more suitable for ASR. It is difficult to set the training target and loss for the speech enhancement network, since the ASR preference on the enhanced signal is not very clear. One promising way is to optimize the speech enhancement network directly by the ASR loss, as is done in \cite{xiao2016,li2016neural,meng2017}. One difficulty for this is the joint training of the speech enhancement network and ASR network, which suffers from the local optima problem. To mitigate this problem, the speech enhancement network can be first pre-trained, such as with the target of delay and sum beamformer (DSB) and log-magnitude spectra in \cite{xiao2016}, or with the target of DSB enhanced signal in \cite{meng2017}. In this work, we focus on the narrow-band speech enhancement network itself, and its joint training with ASR network is left for future work. We set the training loss to a regular speech enhancement loss, i.e. the MSE loss of the STFT coefficient $\|\mathbf{s}(t)-\hat{\mathbf{s}}(t)\|^2$ (the one also used for the CC target). In the present spatial filtering framework, this loss is in the same spirit as the multichannel Wiener filter. We refer to this loss simply as spatial filter (SF). 
}

\addnote[spatial3]{1}{
This loss is optimal in the speech enhancement sense, which however is not necessarily optimal for ASR. We think one important characteristic of beamforming that makes it good at ASR is that: the parameters, e.g. RTF and noise covariance matrix, are normally estimated with a long-term temporal smoothing, thence the beamformer is just slowly time-varying as well, which keeps the temporal consistence of both the enhanced speech and the residual noise. In other words, beamforming does not cause abrupt artifacts. Based on this assertion, we revise the training loss to smooth the spatial filter as:
\begin{equation}\label{eq:ssf}
\|\mathbf{s}(t)-\hat{\mathbf{s}}(t)\|^2 + \lambda \|\mathbf{w}(t)-{\mathbf{w}}(t-1)\|^2
\end{equation}  
where $\lambda$ denotes the weight for the smoothing loss. This loss will be referred to as smoothed spatial filter (SSF).
}

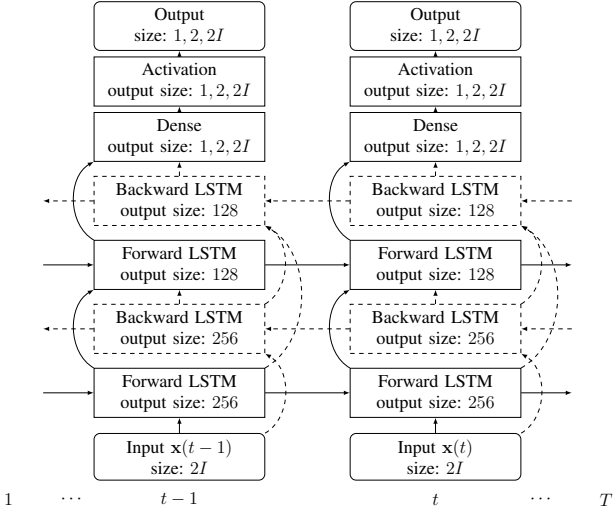
\begin{figure}[t!]
\centering
\large
\pgfdeclarelayer{background}

\pgfsetlayers{background}

\usetikzlibrary{shapes.geometric,backgrounds,arrows,calc,fit}
\tikzstyle{io} = [rectangle, rounded corners, minimum width=4cm, minimum height=1.0cm, text centered, draw=black,align=center]
\tikzstyle{lstm} = [rectangle, minimum width=4cm, minimum height=0.5cm, text centered, draw=black,align=center]
\tikzstyle{blstm} = [rectangle,dashed, minimum width=4cm, minimum height=0.5cm, text centered, draw=black,align=center]
\tikzstyle{data1} = [rectangle, rounded corners, minimum width=4cm, minimum height=0.8cm, text centered, draw=black,align=center]
\tikzstyle{nobox} = [rectangle, rounded corners, draw=none,fill=white,minimum width=1cm, minimum height=0cm, text centered, align=center]
 \tikzstyle{constant} = [rectangle, minimum width=4cm, minimum height=1.1cm,text centered, draw=black,align=center] 
 
 \tikzstyle{arrow} = [->,>=latex]
 \tikzstyle{aarrow} = [->,>=latex,dashed]

\resizebox{0.95\columnwidth}{!}{
\begin{tikzpicture}

%% main flow

\node (xt) [io] {{Input} $\mathbf{x}(t)$ \\ size: $2I$};

\node (t) [below of=xt,node distance=1.0cm] {\large $t$};
\node (d1) [right of=t,node distance=2.5cm] {$\cdots$};
\node (T) [right of=d1,node distance=1.5cm] {$T$};

\node (l1t) [lstm,above of =xt,node distance=1.5cm] {Forward LSTM \\ output size: $256$};
\node (bl1t) [blstm,above of =l1t,node distance=1.5cm] {Backward LSTM \\ output size: $256$};

\node (l2t) [lstm,above of =bl1t,node distance=1.5cm] {Forward LSTM \\ output size: $128$};
\node (bl2t) [blstm,above of =l2t,node distance=1.5cm] {Backward LSTM \\ output size: $128$};

\node (denset) [lstm,above of =bl2t,node distance=1.5cm] {Dense \\ output size: $1, 2, 2I$};
\node (sigt) [lstm,above of =denset,node distance=1.3cm] { Activation \\ output size: $1, 2, 2I$};
\node (yt) [io,above of =sigt,node distance=1.3cm] {{Output} \\ size: $1,2,2I$};

\node (xtm1) [io,left of =xt,node distance=6.0cm] {{Input} $\mathbf{x}(t-1)$ \\ size: $2I$};

\node (tm1) [below of=xtm1,node distance=1.0cm] {\large $t-1$};
\node (d2) [left of=tm1,node distance=2.5cm] {$\cdots$};
\node (t1) [left of=d2,node distance=1.5cm] {$1$};

\node (l1tm1) [lstm,above of =xtm1,node distance=1.5cm] {Forward LSTM \\ output size: $256$};
\node (bl1tm1) [blstm,above of =l1tm1,node distance=1.5cm] {Backward LSTM \\ output size: $256$};

\node (l2tm1) [lstm,above of =bl1tm1,node distance=1.5cm] {Forward LSTM \\ output size: $128$};
\node (bl2tm1) [blstm,above of =l2tm1,node distance=1.5cm] {Backward LSTM \\ output size: $128$};

\node (densetm1) [lstm,above of =bl2tm1,node distance=1.5cm] {Dense \\ output size: $1, 2, 2I$};
\node (sigtm1) [lstm,above of =densetm1,node distance=1.3cm] {Activation \\ output size: $1, 2, 2I$};

\node (ytm1) [io,above of =sigtm1,node distance=1.3cm] {{Output} \\ size: $1, 2, 2I$};

\draw [arrow]  (xt) --  (l1t);
\draw [aarrow] (xt.north east) to [out=30,in=330] (bl1t.south east);
\draw [arrow] (l1t.north west) to [out=150,in=210] (l2t.south west);
\draw [aarrow] (l1t.north east) to [out=30,in=330] (bl2t.south east);
\draw [aarrow] (bl1t) --  (l2t);
\draw [aarrow] (bl1t.north east) to [out=30,in=330] (bl2t.south east);
\draw [arrow] (l2t.north west) to [out=150,in=210] (denset.south west);
\draw [aarrow] (bl2t) --  (denset);
\draw [arrow] (denset) --  (sigt);
\draw [arrow] (sigt) --  (yt);

\draw [arrow]  (xtm1) --  (l1tm1);
\draw [aarrow] (xtm1.north east) to [out=30,in=330] (bl1tm1.south east);
\draw [arrow] (l1tm1.north west) to [out=150,in=210] (l2tm1.south west);
\draw [aarrow] (l1tm1.north east) to [out=30,in=330] (bl2tm1.south east);
\draw [aarrow] (bl1tm1) --  (l2tm1);
\draw [aarrow] (bl1tm1.north east) to [out=30,in=330] (bl2tm1.south east);
\draw [arrow] (l2tm1.north west) to [out=150,in=210] (densetm1.south west);
\draw [aarrow] (bl2tm1) --  (densetm1);
\draw [arrow] (densetm1) --  (sigtm1);
\draw [arrow] (sigtm1) --  (ytm1);

\draw [arrow] (l1tm1) --  (l1t);
\draw [arrow] (l2tm1) --  (l2t);
\draw [aarrow] (bl1t) --  (bl1tm1);
\draw [aarrow] (bl2t) --  (bl2tm1);

\node (l2tm2) [nobox, left of =l2tm1,node distance=3.7cm] {};
\draw [arrow]  (l2tm2) --  (l2tm1);
\node (l1tm2) [nobox, left of =l1tm1,node distance=3.7cm] {};
\draw [arrow]  (l1tm2) --  (l1tm1);
\node (l2tp1) [nobox, right of =l2t,node distance=3.7cm] {};
\draw [arrow]  (l2t) --  (l2tp1);
\node (l1tp1) [nobox, right of =l1t,node distance=3.7cm] {};
\draw [arrow]  (l1t) --  (l1tp1);

\node (bl2tm2) [nobox, left of =bl2tm1,node distance=3.7cm] {};
\draw [aarrow]  (bl2tm1) --  (bl2tm2);
\node (bl1tm2) [nobox, left of =bl1tm1,node distance=3.7cm] {};
\draw [aarrow]  (bl1tm1) --  (bl1tm2);
\node (bl2tp1) [nobox, right of =bl2t,node distance=3.7cm] {};
\draw [aarrow]  (bl2tp1) --  (bl2t);
\node (bl1tp1) [nobox, right of =bl1t,node distance=3.7cm] {};
\draw [aarrow]  (bl1tp1) --  (bl1t);

\end{tikzpicture}
}
\vspace{-0.0cm}
\caption{Diagram of the proposed architecture. The unidirectional (forward) LSTM is represented with solid-lines blocks and arrows, while the additional blocks and arrows needed for BLSTM are represented with dashed lines.} 
\label{fig:lstm}
\vspace{-0.0cm}
\end{figure}

%\end{document}%

\subsection{Network Architectures}

%RNN transmits the hidden units along time step. To avoid the problem of exponential weight decay (or explosion) along time steps, LSTM introduces an extra memory cell, which conveys the information along time step respectively to the hidden units. The memory cell allows to learn long-term dependencies. For the detailed structure of LSTM, see the seminal paper \cite{hochreiter1997}.

The architectures of the proposed LSTM and bidirectional LSTM (BLSTM) networks are shown on Fig. \ref{fig:lstm}. It maps the input sequence onto the output sequence. Two LSTM layers are stacked. Through a dense layer, the output vector of the second LSTM layer is mapped onto the output vector. Then an activation is applied to obtain the network output. The output size of LSTM layers are set based on preliminary experiments. Notice that this figure summarizes three networks with three different targets and associated outputs, namely MRM, CC, and SF. While the input sequence at frequency bin $k$ is the same for all three networks, namely $\mathbf{X}(k)$ defined in \eqref{eq:xseqnor}, the network outputs and the output dimensions are different. The output sequences $\mathbf{M}(k)$, $\mathbf{S}(k)$ and $\mathbf{W}(k)$, defined by \eqref{eq:mseq}, \eqref{eq:sseq} and \eqref{eq:wseq}, are of dimension 1, 2, and $2I$, respectively. 

Moreover, we chose different activation functions for each one of these networks, namely \emph{sigmoid}, \emph{identity} and \emph{tanh}, respectively. We remind that the same network (same parameters) is trained for all the frequency bins $k\in\{0 \dots K-1\}$.
The number of parameters to be learned slightly varies with the number of microphones and with the dimension of the output. On an average, the LSTM and BLSTM networks have $470,000$ and $1,200,000$ parameters, respectively.

\section{Experiments}
\label{sec:experiment}
\subsection{Experimental Setup}
\label{sec:exp-setup}

\setlength{\tabcolsep}{10pt}
\begin{table*}[t]
\centering
\caption{Network summary of WB-CRNN-MRM, WB-BLSTM1-SF, WB-BLSTM2-SF and the proposed BLSTM-SF, for the 4CH case. }
\label{tab:networks}
\begin{tabular}{c  | c | c | c | c   }  
  &WB-CRNN-MRM \cite{chakrabarty2019} & WB-BLSTM1-SF \cite{meng2017} & WB-BLSTM2-SF \cite{meng2017}  & BLSTM-SF (prop.) \\ \hline
Input dimension    & $4\times 129 \times 2$  & 2056 & 2056 & 8 \\
Network  & 3 CNN ($2\times 1$, 64, out: 8259)  & 1 BLSTM (256) & 2 BLSTM (1024) & 1 BLSTM (256) \\  
         & 1 BLSTM (128) &  1 BLSTM (128) & 1 Dense (2056) & 1 BLSTM (128) \\
         & 1 Dense (512) + 1 Dense (129) &   1 Dense (2056) & & 1 Dense (8)\\
Output dimension  & 129   & 2056 & 2056 & 8 \\
\# Parameters  & 8.8 M & 5.9 M & 54.6 M & 1.2 M \\
Training data  & 19 hours & 11 hours & 56 hours & 11 hours\\
\end{tabular}
\vspace{-.0cm}
\end{table*}

\subsubsection{Data Generation}\label{sec:data}
We use the CHiME4 dataset \cite{barker2015}, which was recorded with six microphones embedded in a tablet device.  CHiME4 toolkit provides a method to simulate the multichannel data. However, instead of using the multichannel frequency responses, this method only simulates the multichannel time delays. Our preliminary experiments show that training the network with this type of simulated data performs poorly with real test data. Therefore, we use real data both for training and for testing purposes. The noise-free multichannel speech data were recorded in a booth (BTH) and the training, development and evaluation data were recorded by three different groups of four speakers.  The multichannel background noise were recorded with four noisy environments, namely bus (BUS), cafe (CAF), pedestrian area (PED), and street junction (STR).  For each type of noise, four to five sessions were recorded at different times, with a duration of about 0.5 hours per session. 

The four speakers in BTH training set (399 utterances) are used for network training, and the eight speakers in BTH development (410 utterances) and evaluation (330 utterances) sets are used for test.
%The first group  (four speakers) are used for training (399 utterances), and the other two groups (eight speakers) are used for development (410 utterances) and for evaluation (330 utterances).
 Each noise session is split into two sub-sessions used for training (60\%) and for test (40\%), respectively, which means that different noise instances are used for training and for test. To generate the training data, noise segments randomly extracted from the training sub-sessions are mixed with BTH training utterances, with signal-to-noise-ratios (SNRs) randomly selected from the interval$[-5,10]$~dB. Each training utterance is mixed with fifteen different randomly selected noise segments, and a total of about 11 hours of training data are generated. 

Two groups of data are tested, (i)~MIXED data: background noise segments randomly extracted from the test sub-sessions are mixed with BTH test utterances, with SNRs in $\{-4, 0, 4, 8\}$ dB.  For each noise type and SNR, about 200 test utterances are generated; (ii)~REAL data: the development (Dev) and evaluation (Eval) sets from CHiME4 real data were recorded in the four noisy locations by the same speakers in both development and evaluation BTH sets. 
 
The signals are transformed to the STFT domain using a 512-sample (32~ms) Hanning window with a frame step of 256 samples. The sequence length for training is set to $T=192$ frames (about 3~s), which means the LSTM network is trained to learn 192 time steps of memory. The training sequences are picked out from the utterance-level signals with 50\% overlap for two adjacent sequences. In total, about 6.55 millions of training sequences are generated.
For test, the utterances are not cut into sequences with length of $192$ frames but, instead, the entire utterances are directly used for sequence-to-sequence prediction. 

\subsubsection{Training Configuration}

We found that the microphone \#1 recording in the evaluation set has a much larger volume than the volume used in other recording sets. The issue of microphone array mismatch is beyond the scope of this work, thus microphone \#1 is not used.  
Microphone \#2 is not used as well, due to its low availability. We conducted experiments with two microphone configurations, i.e. microphones \#3, \#4, \#5 and \#6 (4CH), and microphones \#5 and \#6 (2CH). Microphone \#6 is taken as the reference channel. The network variants are named based on the network type, i.e. LSTM or BLSTM, on the output target, i.e. MRM, CC, SF or SSF. For example, BLSTM-SSF refers to BLSTM with smoothed spatial filter as target. Based on some preliminary experiments, we set the weight $\lambda$ in (\ref{eq:ssf}) for BLSTM-SSF to 1. All these network variants are trained individually from scratch.

We use the Keras environment \cite{chollet2015keras}  to implement the proposed architectures and associated methods. The Adam optimizer \cite{kingma2014adam} is used with a learning rate of 0.001. The batch size is set to 512. The training sequences were shuffled. Based on some preliminary experiments, all the networks are trained with ten epochs. 

\subsubsection{Performance Metrics}
To evaluate and benchmark the speech enhancement performance for the MIXED data, three intrusive metrics are used, (i)~the perceptual evaluation of speech quality (PESQ) \cite{rix2001}  which evaluates the quality of the enhanced signal in terms of both noise reduction and speech distortion,  (ii)~the short-time objective intelligibility (STOI) \cite{taal2011}, a metric that highly correlates with noisy speech intelligibility; and (iii)
the signal-to-distortion ratio (SDR) \cite{vincent2006} in dB measures the level of noise reduction. For all the metrics, the larger the better. The BTH clean signal is taken as the reference signal. 

For REAL data, these intrusive metrics are not used because the close-talk signals provided in the CHiME4 dataset are not reliable. Instead, a non-intrusive metric is used to measure the speech enhancement performance, i.e. the normalized speech-to-reverberation modulation energy ratio (SRMR) \cite{santos2014improved}, which measures the amount of noise, and also reflects the speech intelligibility. In addition, we tested the performance of automatic speech recognition (ASR) obtained with the enhanced signals. The ASR of \cite{hori2015}, with already-trained ASR models and decoding recipe provided in CHiME4 is taken as the baseline system.\footnote{ http://spandh.dcs.shef.ac.uk/chime\_challenge/chime2016/download.html} This system uses mel-frequency cepstral coefficients (MFCC), a DNN-HMM acoustic model and an RNN language model. The DNN-HMM acoustic model is trained using the single-channel noisy multi-condition CHiME4 training data. The ASR performance is measured with the word error rate (WER), the lower the better.

\subsubsection{Comparison Methods}
We compare with the following four multichannel speech enhancement methods:
\begin{itemize}
\item The BeamformIt method of \cite{anguera2007}, based on an unsupervised filter-and-sum beamforming technique;
\item The neural-network based generalized eigenvalue beamformer (NN-GEV) method of \cite{heymann2016}.  A BLSTM network is used to estimate a spectral mask, based on which a generalized-eigenvalue beamformer is computed and applied to speech denoising. We use the toolkit provided by the authors of \cite{heymann2016},\footnote{https://github.com/fgnt/nn-gev} in which the BLSTM parameters had already been trained using the CHiME4 training dataset;
\addnote[wbmethods]{1}{\item The CRNN method of \cite{chakrabarty2019} takes as input multichannel full-band STFT coefficients and predicts single-channel full-band MRM, i.e. \eqref{eq:mrm}. Several CNN layers are employed for each STFT frame to extract the inter-channel information, then followed by one BLSTM layer to learn the inter-frame information, where two past frames and two future frames are taken as the context for each frame. Since the authors' implementation is not publicly available, we implemented the method and used the CHiME4 dataset to train and evaluate \cite{chakrabarty2019}. About 19 hours of training data were used, from which 9.14 millions of training samples were generated. The STFT is conducted with 256-sample frame length and 128-sample frame step. We refer to this method as wide-band CRNN-MRM (WB-CRNN-MRM);
\item The full-band deep beamforming network of \cite{meng2017}. A BLSTM network takes as input the multichannel full-band (real and image parts of) STFT coefficients with dimension of $2KI$, and predicts the full-band (real and image parts of) beamformer with the same dimension as input. To have a fair comparison, we don't follow the training setup presented in \cite{meng2017} -- that trains the speech enhancement network with an ASR loss. Instead, we follow the way presented in this work that uses the MSE loss of the enhanced STFT coefficients. The same STFT configuration with the proposed method is taken. We implemented two networks: (i)~the one presented in Fig.~\ref{fig:lstm}, namely two BLSTM layers are stacked, each layer has 256 and 128 hidden units, respectively. This network is trained using the same amount of data with the proposed method, i.e. about 11 hours.   The training sequences with $192$ frames are picked out from the utterance-level signals with 50\% overlap for two adjacent sequences. This generates about $25,000$ training sequences. This network is referred to as WB-BLSTM1-SF; (ii)~the previous network is obviously too small relative to the input/output dimension. We trained a more appropriate network with two layers stacked, and with 1024 hidden units per layer. About 56 hours of training data were used. This network is referred to as WB-BLSTM2-SF. Both networks are trained with a batch size of 32.}
\end{itemize}

Table \ref{tab:networks} briefly summarizes the four networks, i.e. WB-CRNN-MRM,  WB-BLSTM1-SF, WB-BLSTM2-SF and the proposed BLSTM-SF. Note that the proposed networks with other targets are very close to BLSTM-SF only with a slight difference due to the different output dimensions. One important characteristic of the proposed narrow-band network is the small network size and the low training data demand.

\subsection{Evaluation of Generalization Capability}

The default training setup presented in Section \ref{sec:data} is \textit{speaker independent and noise-type dependent} (SID-ND): even though training and test use different noise instances, they both use all the four noise types.  
To evaluate the generalization capability in terms of speaker identity and of noise type, two extra training setups are also tested 
%in order to evaluate the generalization capabilities of the proposed model
: (i)~
\textit{speaker independent and noise-type independent} (SID-NID): four speakers are used for training and the other eight speakers are used for test, and three noise types are used for training and the other noise type is used for test, and (ii) \textit{speaker dependent and noise-type dependent} (SD-ND): all twelve speakers and all four noise types are used to generate training data. For each method, a similar amount of training data were generated for all these three configurations. 

\begin{figure}[t!]
\centering
\includegraphics[width=.98\columnwidth]{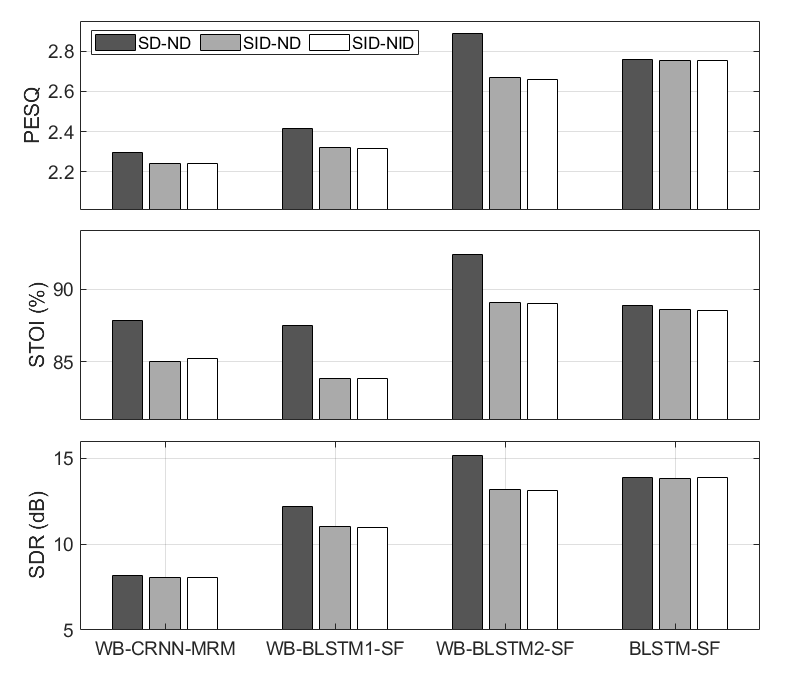}
\caption{\small{Speech enhancement results for the MIXED data with three different training configurations, \emph{speaker independent and noise-type dependent} (SID-ND),  \emph{speaker independent and noise-type independent} (SID-NID), and  \emph{speaker dependent and noise-type dependent} (SD-ND). These results are the averaged scores over the four environments, with a SNR of 0~dB, and for the 4CH case. The PESQ, STOI and SDR of the unprocessed signals are 1.60, 73.8\% and 0.4 dB, respectively.}}
\label{fig:mixed-generalization}
\end{figure}

Fig.~\ref{fig:mixed-generalization} shows the speech enhancement results obtained with the MIXED data for these three training configurations. 
 For the wide-band methods, i.e. WB-CRNN-MRM, WB-BLSTM1-SF and WB-BLSTM2-SF, the speaker dependent case, i.e. SD-ND, noticeably outperforms the speaker independent case, i.e. SID-ND.  The noise-type dependent/independent configurations, i.e. SID-ND and SID-NID, achieve similar performance. The wide-band methods takes as input the full-band multichannel STFT coefficients, which include all the spectral, temporal and spatial informations. Inevitably, the network will learn the spectral pattern of signals, and thus it has the problem to generalize to unseen speakers that have new spectral patterns. These methods generalize well in term of noisy type, possibly since the spectral pattern of each CHiME4 noise type can be well covered by the other three noise types.  
 
We here only show the results for BLSTM-SF, and the proposed network with other targets behave similarly as BLSTM-SF. The proposed narrow-band network achieves comparable performance for all the three configurations, which means it has good generalization capabilities in terms of both speaker and noise type. The network is trained using narrow frequency bands, hence the wide-band spectral-pattern differences between the training and test samples, of both speech and noise, are not taken into account and hence they shouldn't have an impact on the generalization capabilities of the proposed model. The network is actually trained to learn some functions based on the temporal and spatial characteristics of speech and noise, which are independent with respect to their spectral content.
In addition, in the CHiME4 data, the microphone-to-speaker relative positions are time-varying for both the training and test data, which means that the proposed method also generalizes well in terms of moving speakers. Overall, the proposed model learn features that are suitable across frequency bins, as well as for unseen speakers and noise types. 

The wide-band methods have the speaker generalization problem when using only four training speakers, which can be mitigated by increasing the number of training speakers. To fully compare the speech enhancement capabilities regardless of speaker generalization, we report the SD-ND results for the wide-band methods in the following experiments.

\setlength{\tabcolsep}{5pt}
\begin{table*}[t]
\centering
\caption{Speech enhancement results obtained with the MIXED data. SNR is of 0 dB.}
\label{tab:mixed_result}
\begin{tabular}{c l | c c c c c | c c c c c | c c c c c }   
   &    & \multicolumn{5}{c|}{PESQ $\uparrow$}  &  \multicolumn{5}{c|}{STOI (\%) $\uparrow$} &  \multicolumn{5}{c}{SDR (dB) $\uparrow$} \\  
	&   & BUS & CAF & PED & STR & Average & BUS & CAF & PED & STR & Average & BUS & CAF & PED & STR & Average \\ \hline
& unproc.   &  1.93 &   1.47 &   1.43 &   1.57  &  1.60  & 
82.6 &    69.5 &   67.2  &  76.0  &  73.8 &
0.3  &  0.6 &   0.1  &  0.6  &  0.4 \\ \hline
& BeamformIt  \cite{anguera2007} & 2.03 &    1.55 &    1.51 &    1.66 &    1.69 
& 83.7  & 71.6  &  70.1 &  77.1 &  75.7 &
0.1 &    0.5 &    0.5 &    0.4 &    0.4   \\
& NN-GEV  \cite{heymann2016}& 2.12 &    1.57 &    1.61 &    1.76 &    1.77 &
86.7 &    75.1 &    74.9 &    81.8 &    79.6  &
1.8 &    1.9 &    2.1 &    2.3 &    2.0  \\
& WB-CRNN-MRM \cite{chakrabarty2019} & 2.59 &    1.88 &    1.80 &    2.14 &    2.10   &
89.6 &  81.4  &    79.6  &   86.0 &    84.2 &
9.6 &    6.8 &    5.6 &    8.1 &    7.5  \\
& WB-BLSTM1-SF \cite{meng2017} & 2.80 &    1.99 &    1.96 &    2.38 &    2.28  &
90.7 &   80.8 &   79.9 &   86.8 &   84.6 &
 13.7 &  9.0 &    8.3  &    11.3 &    10.6    \\
& WB-BLSTM2-SF \cite{meng2017} & 3.22 &    2.41 &    2.39 &    2.84 &    2.72  &
94.7 &   87.8 &   87.2 &   92.2 &   90.4 &
16.9 &   11.8 &   11.0 &   14.0 &   13.4   \\
2CH & BLSTM-MRM & 2.85 &    2.15 &    2.08 &    2.44 &    2.38  &
89.4 &  80.1 &   78.3 &   85.1 &   83.2  &
12.5 &    9.2 &    8.1 &    10.4 &   10.1   \\
& BLSTM-CC & 2.92 &   {2.14}  &   2.10 &    {2.48} &    2.41 &
{90.5} &  80.4 &   {78.9} &   {85.8} &   {83.9}  &
14.1 &    10.0 &   9.4 &   11.6 &   11.3   \\  
& BLSTM-SF  &  {2.93} &   {2.15} &    {2.11} &    {2.49}  &   {2.42} & 
90.4 &   80.4&   79.0 &   {85.9} &   83.9 &
14.3 &   10.0 &  9.5 &    11.7 &   11.4    \\ 
& BLSTM-SSF  &  2.62 &  2.00 &   1.91 &   2.23  &   2.19 & 
88.9 &   79.5&   77.5 &  84.4 &   82.6 &
12.4 &   9.6 & 8.6 &  10.7 &   10.3    \\\hline
& BeamformIt  \cite{anguera2007} & 2.07 &   1.60  &  1.56 &   1.68 &  1.72 &
85.0 &   74.2 &  72.5 &   78.1 &   77.4 &
0.4 &   0.5 &    1.0 &    0.2 &    0.5  \\
& NN-GEV  \cite{heymann2016} & 2.37 &    1.77 &    1.79 &    2.00 &    1.98 &   
90.6  &  83.3 &   82.9  &   89.0 &  86.4&
3.6 &    4.2 &    4.8 &    4.4  &    4.3   \\
& WB-CRNN-MRM \cite{chakrabarty2019} & 2.77 &    2.11 &    1.97 &    2.34 &    2.30 &
91.4 &   86.4 &   84.5 &   89.1 &   87.9 &
8.0 &    9.7 &    7.6 &    6.4 &    8.4  \\
& WB-BLSTM1-SF \cite{meng2017} & 2.92 &  2.15 &  2.10 &  2.49 &   2.41  &
92.3 &   84.8 &  83.7 &  89.1 &   87.5 &
15.0 &   10.7 &  10.2 &  12.9 &  12.2   \\
& WB-BLSTM2-SF \cite{meng2017} & 3.39 &   2.60 &    2.56 &    3.00 &    2.89  &
95.7 &  90.4 &  89.6 &   93.8 &   92.4 &
18.3 &  13.6 &  12.8 &   16.0 &   15.2   \\
4CH & BLSTM-MRM & 3.10 &    2.42 &    2.32 &    2.71 &    2.64 &
91.3 &  85.3 &  83.0  & 88.8  & 87.1 &
13.1 &  10.2 &  9.1 &  11.1 &   10.9  \\
& BLSTM-CC & 3.28 &    {2.53} &   {2.41} &    2.87 &    2.77 &
{92.3}  & 86.2  & 83.9  & {90.5}  & {88.3} &
16.6 &   13.0 &   12.0 &   14.9 &   14.1  \\
& LSTM-SF   & 3.01 &    2.28 &    2.17 &    2.63 &    2.52 & 
90.7 &    83.1 &    81.0 &    88.4 &    85.8 &
14.5 &   10.9 &    10.0 &   12.9 &   12.1  \\
& BLSTM-SF  &  3.23 &    {2.52} &    {2.41} &    {2.85} &   {2.76} & 
91.7  & {85.6}  & {83.4} &  89.7  & 88.6 &
16.1  &  12.8  &  11.8 &   14.9  &  13.9              \\ 
& BLSTM-SSF  &  3.04 &   2.40 &   2.28 &    2.66 &    2.60 & 
91.7  & 85.6  & 83.4 &  89.7  & 87.6 &
15.1  & 12.2  & 11.3 & 13.8  & 13.1            
\end{tabular}
\end{table*}  

\subsection{Unidirectional versus Bidirectional LSTM}

\begin{figure}[t!]
\centering
\includegraphics[width=.90\columnwidth]{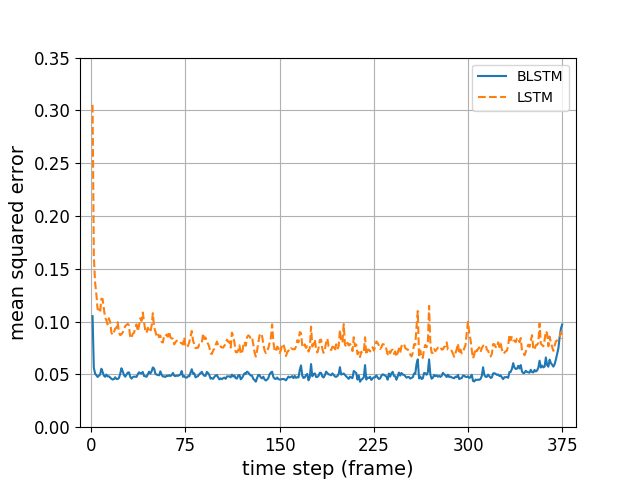}
\caption{\small{The loss evolution, i.e. MSE, as a function of time step, for the proposed BLSTM-SF (blue) and LSTM-SF (orange) methods (with 4CH).}}
\label{fig:mse}
\end{figure}

As presented in Section \ref{sec:exp-setup}, we perform sequence-to-sequence network training using fixed-length sequences with $T=192$ frames, which means the back propagation (through time) of gradients is truncated at 192 time steps. In other words, the network is trained to learn 192 time steps of memory. At test, the network predicts length-varying utterances. Utterances with different lengths have different memory lengths, moreover, different time steps in one utterance have differrent forward/backward memories. To analyze how the memories work, and how many time steps could be memorized in the proposed narrow-band LSTM framework,  Fig. \ref{fig:mse} shows the MSEs as a function of time step. To obtain this plot, we generated one extra group of data (we used the same data generation protocol as with the MIXED test data), which includes 1.3 million sequences with a fixed length of $T=375$ frames (six seconds). The MSEs averaged over all the sequences are shown in Fig. \ref{fig:mse}. The MSE of LSTM  quickly drops from 0.3 to 0.1 in a few time steps, which means a few past frames are already very effective to reduce the loss. The MSE of LSTM then slowly converges to 0.077 in about 150 time steps, which means that, for one time step, the frames earlier than about 150 time steps do not contribute anymore. This is due to one of the following reasons (or the combination of them): (i)~the LSTM network is only able to learn the memory of about 150 time steps, and (ii)~about 150 time steps already provide enough context information in terms of the temporal and spatial properties of the signal.  

When future frames are used, the MSE drops from 0.077 for LSTM to about 0.05 for BLSTM. At the two ends, BLSTM has a larger MSE due to the insufficient past or future context. At the end part, BLSTM has enough past context. The MSE is reduced from 0.097 at the 375-th frame to 0.06 at the 369-th frame, and to 0.05 at the 350-th frame. This indicates that, when enough past context is being used, about six future frames are already very effective to reduce the loss, and about 25 future frames provide sufficient information to further reduce the loss to a satisfactory value. For an online application, past information is always available. The amount of future frames to be used can be chosen as a trade-off between performance and processing latency: (i)~25 future frames can be used to have the best prediction performance that BLSTM can achieve, which however leads to a 400~ms latency, (ii)~6 future frames can be chosen to have a good performance with 96~ms latency, which is not a problem from a practical point of view.

Tables \ref{tab:mixed_result} and \ref{tab:real_result} show the experimental results obtained with the MIXED and REAL data, respectively. Comparing the results of LSTM-SF and of BLSTM-SF, one can see that BLSTM performs, indeed, noticeably better than LSTM in terms of both speech enhancement and speech recognition. A larger error obtained with LSTM than with BLSTM would lead to a larger speech distortion and to less noise reduction. The difference in performance between LSTM and BLSTM can easily be perceived by listening to the enhanced signals.
The comparison between LSTM and BLSTM, based on the performance of LSTM-SF and BLSTM-SF (with 4CH), also holds for other proposed targets and numbers of channels. In the following, we will only analyze the performance of BLSTM networks.

\subsection{Speech Enhancement Results with MIXED Data}

Table \ref{tab:mixed_result} shows the speech enhancement results obtained with the MIXED data and with an SNR of 0~dB. It is not surprising that the 4CH cases outperform the 2CH ones, due to the use of richer spatial information. In the following, we will mainly compare the 4CH performance scores (the comparison is equally valid for the 2CH cases). 

Over the unprocessed signals, BeamformIt improves the scores to a certain extent. NN-GEV, which uses a deep neural network to classify the speech and noise TF bins, performs much better than Beamformit. It was demonstrated in \cite{heymann2016} that the speech enhancement performance of NN-GEV is quite close to the performance of an oracle beamformer, while the oracle one uses the true speech/noise  classification for the beamformer estimation. 
All the other methods prominently outperforms these two beamformers. This indicates that the techniques that directly predict the clean speech with neural network have a better noise reduction capability than the beamforming techniques. 
 
WB-CRNN-MRM and the proposed BLSTM-MRM both predict the magnitude ratio masks (MRMs), while the latter  achieves similar STOI scores and better PESQ and SDR scores. Better PESQ and SDR scores indicate more noise reduction. In \cite{chakrabarty2019}, it was stated that WB-CRNN-MRM mainly exploits the wide-band spatial characteristics to distinguish between speech and noise, by first extracting the inter-channel (spatial) information with CNNs and then exploiting its short-term (five frames) temporal dynamics with BLSTM. We can see from this comparison that, compared to using the wide-band spatial information, fully exploiting the narrow-band temporal-spatial information is more powerful for speech/noise discrimination. 

WB-BLSTM1-SF, WB-BLSTM2-SF and the proposed BLSTM-SF all predict a spatial filter and minimize the MSE loss of the STFT coefficients. WB-BLSTM1-SF (and WB-BLSTM2-SF) takes as input the full-band STFT coefficients of the multichannel noisy signals, which attempt to fully exploit temporal-spectral-spatial information. This wide-band method requires a big network and a large amount of training data to tackle the very high input/output dimensions, as WB-BLSTM2-SF (with 54.6 M parameters and 56 hours of training data) achieves far better performance measures than WB-BLSTM1-SF (with 5.9 M parameters and 11 hours of training data). With the similar networks and the same amount of training data, the proposed BLSTM-SF noticeably outperforms WB-BLSTM1-SF, since BLSTM-SF adequately learns the narrow-band information. When the wide-band network is well trained, compared to BLSTM-SF, WB-BLSTM2-SF indeed achieves higher performance measures, but at the cost of using a much larger network and more training data, and the cost of suffering the speaker generalization problem. 

BLSTM-CC and BLSTM-SF both target the STFT coefficients, and achieve comparable speech enhancement performance. This indicates that the use of spatial filter does not have a significant impact on speech enhancement performance. BLSTM-CC (and BLSTM-SF) improves the performance over BLSTM-MRM by recovering the complex spectra of clean speech. As already mentioned in Section \ref{sec:cc}, to estimate the complex spectra of clean speech in narrow-band, what we expect to use is the spatial features of signals. This expectation is verified by the experimental results: when using more microphones, the prediction of the complex spectra is more reliable, and correspondingly the superiority of BLSTM-CC over BLSTM-MRM is more prominent. For example, for the 2CH case, the PESQ (resp. SDR) score of BLSTM-MRM and BLSTM-CC are 2.38 (resp. 10.1 dB) versus 2.41 (11.3 dB), while these scores for the 4CH case are 2.64 (resp. 10.9 dB) versus 2.77 (resp. 14.1 dB). BLSTM-SSF smooths the spatial filter to keep the temporal consistence of the enhanced signal, which however violates the optimal estimation of clean speech. As a result, the speech enhancement scores are degraded relative to the ones of BLSTM-SF.

\begin{figure*}[t]
\centering
\subfloat[clean speech]{\includegraphics[width=0.32\textwidth]{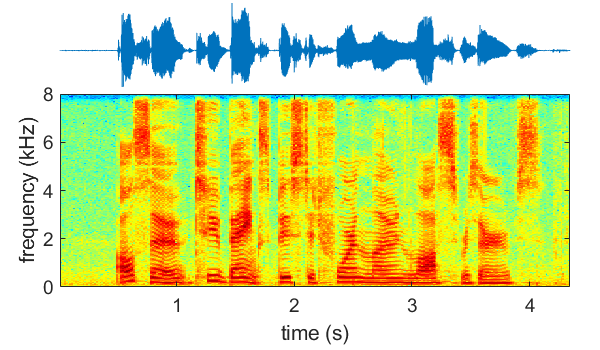}} 
\subfloat[noisy (unproc.)]{\includegraphics[width=0.32\textwidth]{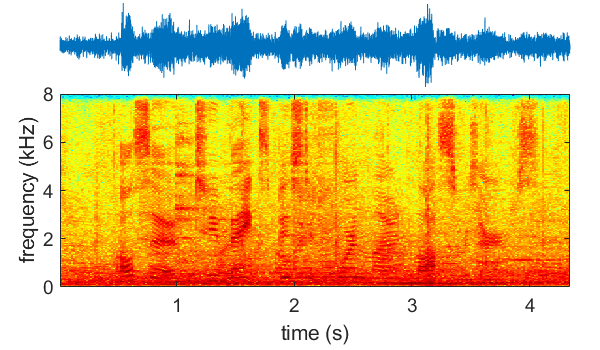}} \\ 
\subfloat[BeamformIt]{\includegraphics[width=0.32\textwidth]{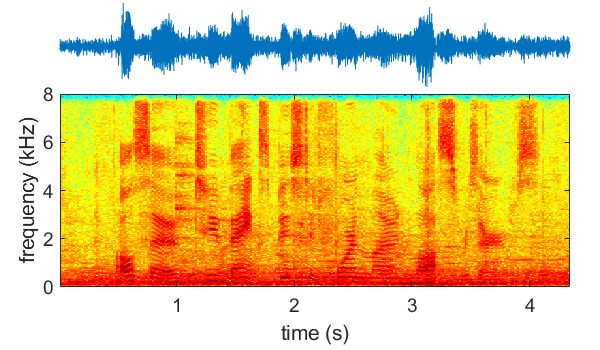}} 
\subfloat[NN-GEV]{\includegraphics[width=0.32\textwidth]{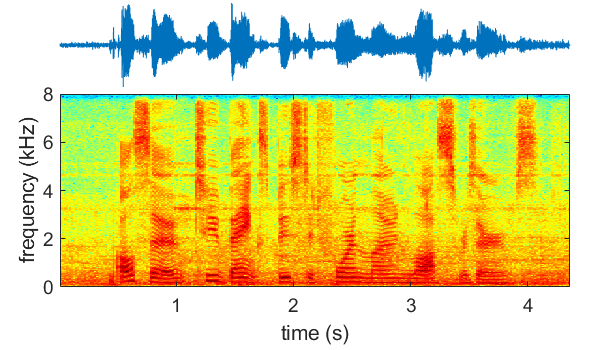}} 
\subfloat[WB-CRNN-MRM]{\includegraphics[width=0.32\textwidth]{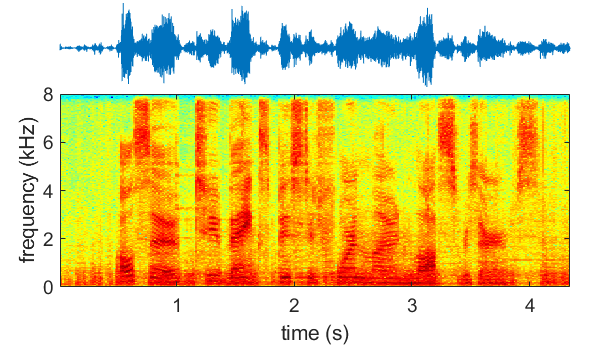}} \\
\subfloat[WB-BLSTM1-SF]{\includegraphics[width=0.32\textwidth]{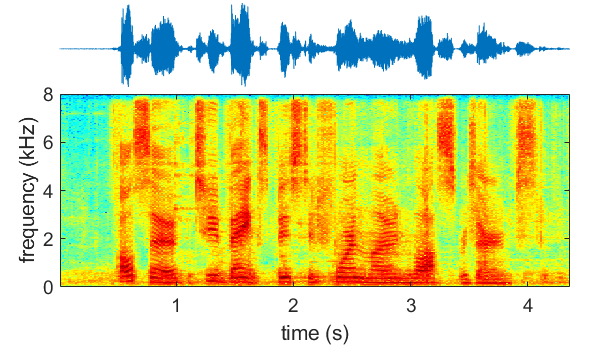}} 
\subfloat[WB-BLSTM2-SF]{\includegraphics[width=0.32\textwidth]{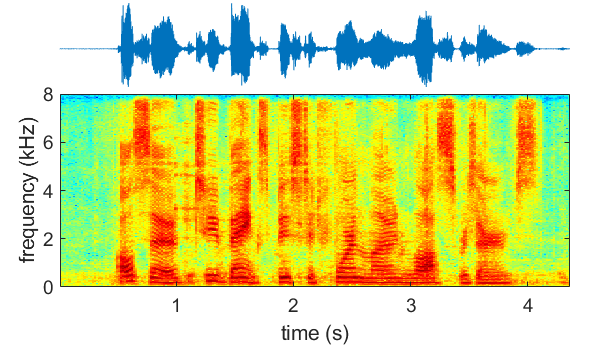}} 
\subfloat[BLSTM-MRM]{\includegraphics[width=0.32\textwidth]{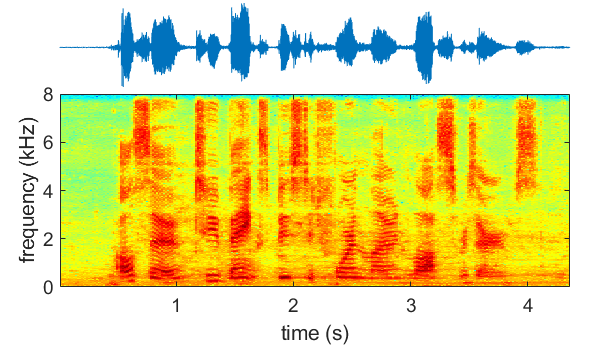}} \\
\subfloat[BLSTM-CC]{\includegraphics[width=0.32\textwidth]{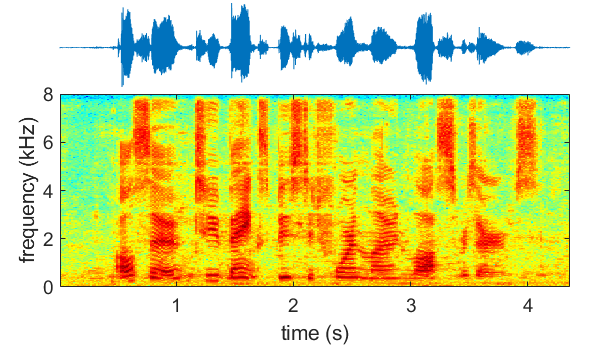}} 
\subfloat[BLSTM-SF]{\includegraphics[width=0.32\textwidth]{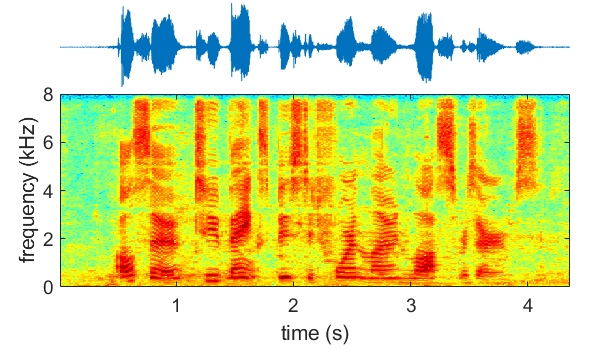}} 
\subfloat[BLSTM-SSF]{\includegraphics[width=0.32\textwidth]{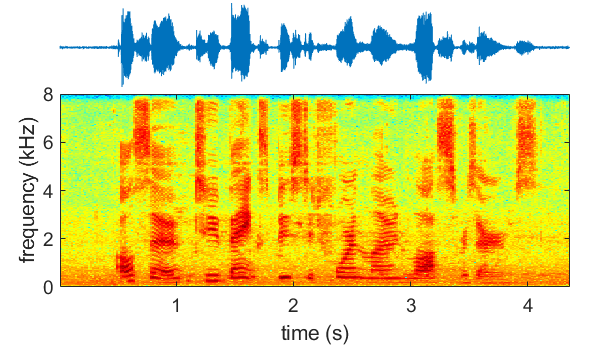}} \\
\caption{\small Waveforms and spectrograms of the clean-speech input, of the added noise and of the results obtained with state of the art methods and with the proposed BLSTM models, associated with one utterance from the MIXED dataset using four channels (4CH). In this example, CAF noise is added to the clean speech signal and the SNR is of 0~DB.}  
\label{fig:example}
\end{figure*} 

Fig. \ref{fig:example} shows waveforms and spectrograms associated with one example. It can be seen that two beamformers (Fig. \ref{fig:example} (c) and (d)) well preserve the speech spectra, while a large amount of noise still remain, which corresponds to the low speech enhancement scores presented in Table \ref{tab:mixed_result}. Three wide-band methods, i.e. WB-CRNN-MRM, WB-BLSTM1-SF and WB-BLSTM2-SF (Fig. \ref{fig:example} (e), (f) and (g)) largely remove the noise and recover the speech structure. However, the recovered speech spectra look somewhat blurred along the frequency axis, and some wide-band spectra are wrongly deleted or inserted. These types of wide-band prediction error are caused by that: for high-dimensional (full-band) regression, the networks are not fully capable of recovering the details of the high-dimensional output vector, and prediction errors are highly correlated between vector elements (frequencies). The wide-band prediction errors lead to some audible abrupt distortions/interferences by listening to the enhanced signals. In contrast, the proposed narrow-band methods (Fig. \ref{fig:example} (h)-(k)) don't produce the wide-band distortions, due to the untied frequencies. It is consistent to the results of Table \ref{tab:mixed_result} that BLSTM-CC and BLSTM-SF perform similarly, and remove more noise than BLSTM-MRM and BLSTM-SSF. In the very low frequency region, the proposed methods failed to properly predict the speech spectra due to the very low SNR in this region. For this case, the wide-band networks work well by predicting all the frequencies together. 

\begin{figure}[t]
\centering
{\includegraphics[width=0.48\columnwidth]{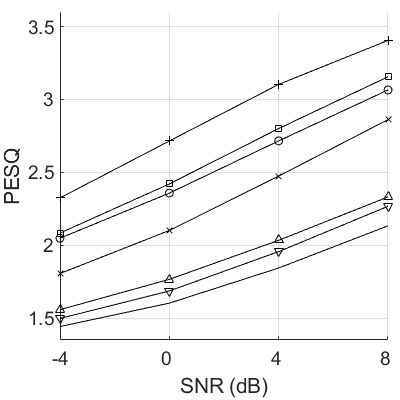}} 
{\includegraphics[width=0.48\columnwidth]{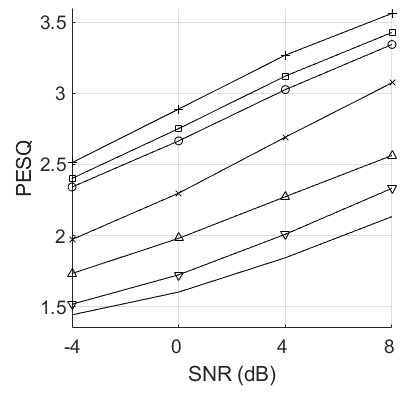}}  \\
{\includegraphics[width=0.48\columnwidth]{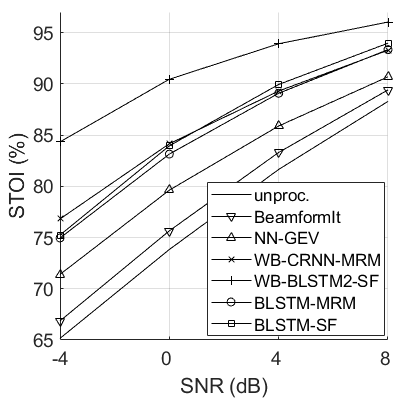}} 
{\includegraphics[width=0.48\columnwidth]{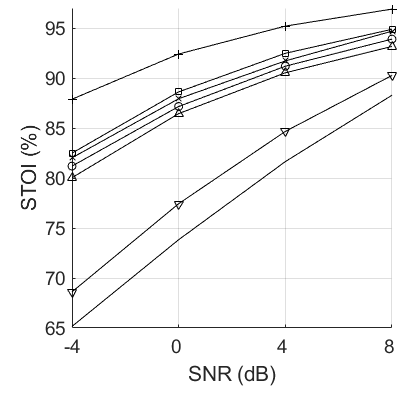}} \\
{\includegraphics[width=0.48\columnwidth]{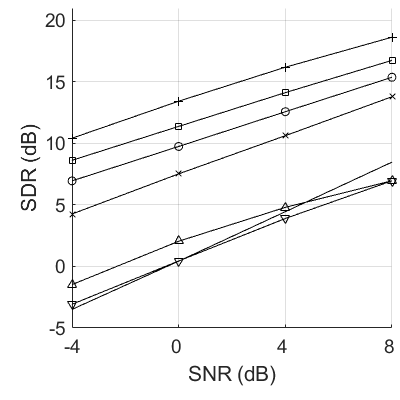}} 
{\includegraphics[width=0.48\columnwidth]{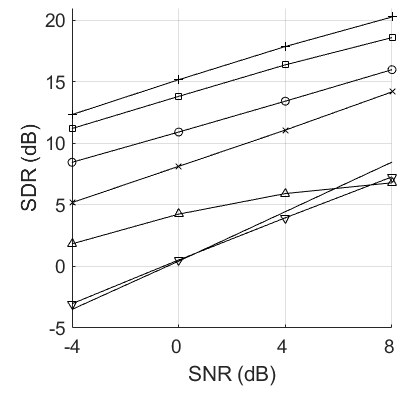}} 
\caption{\small Speech enhancement results obtained with the MIXED data, averaged over all noise types, as a function of SNR, for the 2CH case (left) and the 4CH case (right).  } 
\label{fig:mixed_snr}
\end{figure}

Results obtained with other SNR values are shown in Fig. \ref{fig:mixed_snr}. For the sake of clarity of illustration, the curves of WB-BLSTM1-SF, BLSTM-CC and BLSTM-SSF are not shown. It can be seen that the conclusions drawn above hold for a wide range of SNR values, except that Beamformit and NN-GEV don't improve the SDR of unprocessed signals for the high SNR case.

\subsection{Results with REAL Data}

Table \ref{tab:real_result} shows the speech enhancement and speech recognition scores obtained with the REAL data. The speech enhancement results, i.e. the SRMR scores, are broadly consistent to the results of MIXED data: WB-BLSTM2-SF performs the best, while the proposed methods perform not as good as WB-BLSTM2-SF, but still achieve very high SRMR scores. 

As already mentioned, all the proposed networks were trained with ten epochs, which achieves approximately the best speech enhancement performance, and a few more or less epochs don't lead to a notable performance change. However, this is not true for the ASR performance. For ASR, we take the network that achieves the smallest Dev WER, from the networks trained with 6 to 10 epochs. In addition, the WER performance is not very stable from one trial to another. Therefore, we run five trials for each of the proposed networks, and the averaged scores are reported in Table \ref{tab:real_result}. Even though formal significance test has not been conducted, these WER scores are quite reliable.  

BeamformIt largely reduces the WERs obtained with unprocessed signals.
The ASR performance of the wide-band methods, i.e. WB-CRNN-MRM, WB-BLSTM1-SF and WB-BLSTM2-SF, don't exceed the ones of BeamformIt. WB-BLSTM1-SF performs the worst, and for the 2CH case even degrades the performance of the unprocessed signals. It was demonstrated in \cite{wang2019bridging} that the processing errors of one speech enhancement method have a big impact on the ASR performance. The unsatisfactory ASR performance of the wide-band methods is may caused by their wide-band prediction errors, i.e. the blurred and wrongly deleted/inserted wide-band spectra. This type of wide-band error has not been seen by the ASR training data, and thus causes the mismatch between the training and test data. The WERs of WB-BLSTM2-SF reported here are consistent with the ones presented in \cite{meng2017} that the wide-band deep beamformer does not perform as well as BeamformIt, even when it is jointly trained with the ASR acoustic model.  

NN-GEV and the proposed methods process frequencies independently, and thence significantly outperform the wide-band methods. NN-GEV averagely performs the best for the 4CH case, while performs worse than the proposed methods for the 2CH case.    
The prossible reason for this is: a good beam-pattern of the microphone array is critical for the beamforming techniques, which however requires a large number of microphones. \addnote[mrmasr]{1}{For the 2CH case, BLSTM-MRM performs better than BLSTM-CC and BLSTM-SF, especially achieves the smallest Eval WERs. As already analyzed, the estimation of the clean complex spectra in narrow-band relies on the spatial features of signals, and thus the estimation error is highly related to the number of microphones. When using only two microphones, the high estimation error may degrade the ASR performance. In addition, considering that ASR normally takes the (log-)magnitude feature, the recovery of clean phase may be not really helpful for improving the ASR performance. Even for the 4CH case, BLSTM-MRM achieves a comparable performance as BLSTM-CC.}
For both the 2CH and 4CH cases, BLSTM-SF consistently (across all the conditions) performs better than BLSTM-CC, even though the performance gap is small. This indicates that, to a certain extent ASR indeed benefits from the use of a spatial filter. By further smoothing the spatial filter, BLSTM-SSF notably improves the ASR performance over BLSTM-SF, which verifies our analysis about the beamforming techniques that the temporal smoothness of beamformer is important for its good ASR performance. For the 2CH case, BLSTM-SSF achieves the best Dev WERs, and slightly worse Eval WERs than BLSTM-MRM. For the 4CH Eval data, BLSTM-SSF achieves comparable average WERs with NN-GEV, especially the WER of BUS, CAF and STR of BLSTM-SSF are all smaller than the ones of NN-GEV.
 
\addnote[observations]{1}{From these experiments, regarding ASR, we would like to emphasize the following important observations: (i) for both wide-band and narrow-band methods, one way to improve the ASR performance is to reduce the level of the network prediction error. However, with comparable error levels, the narrow-band processing artifacts is much less harmful for ASR than the wide-band one, and  (ii) the TF masking plus beamforming technique, i.e. NN-GEV, is powerful for ASR. However, one problem for this type of methods is that its maximum ASR performance is actually determined/limited by the performance of oracle beamformers. As demonstrated in \cite{heymann2016} that NN-GEV already performs closely to the oracle beamformer, improving the TF masking performance will no longer lead to ASR improvement. In contrast, the proposed methods directly interface the speech enhancement network and the ASR module, thus the ASR performance can be improved by reinforcing the speech enhancement network, or by performing joint end-to-end training. This means the proposed methods still have a large potential to be explored, which is left for future work.}

\setlength{\tabcolsep}{4.5pt}
\begin{table*}[t!]
\centering
\caption{Speech enhancement and ASR results obtained with the REAL data, where the SRMR scores are averaged over the developement and evaluation datasets.}
\label{tab:real_result}
\begin{tabular}{c l | c c c c c | c c c c c | c c c c c }   
   &    & \multicolumn{5}{c|}{SRMR $\uparrow$}  &  \multicolumn{5}{c|}{WER $\downarrow$ (\%) Dev } &  \multicolumn{5}{c}{WER $\downarrow$ (\%) Eval }  \\  
	&   & BUS & CAF & PED & STR & Average  & BUS & CAF & PED & STR & Average & BUS & CAF & PED & STR & Average \\ \hline
& unproc.   & 1.75 &    2.00 &    2.18 &    1.97 &   1.98  & 
14.77 & 10.74 & 6.83 & 10.54 & 10.72 &
36.08 & 23.35 & 18.24 & 15.37 & 23.26  \\ \hline
& BeamformIt \cite{anguera2007} & 1.74 &    2.10 &    2.24 &    2.04 &   2.03 &
 14.12 & 7.55 & 5.04 & 8.44 & 8.79 &
 25.97 & 16.85 & 14.01 & 12.57 & 17.35  \\
&  NN-GEV \cite{heymann2016} & 2.04 &    2.26 &    2.38 &    2.24 &    2.23 &
11.30 & 6.55 & \textbf{4.71} & 7.52 & 7.52 &
21.20 & 12.94 & 9.92 & 9.39 & 13.36 \\
& WB-CRNN-MRM \cite{chakrabarty2019} & 2.63 &    2.77 &    2.75 &    2.69 &    2.71  &
13.78 & 10.35 & 7.11 & 10.22 & 10.37 &
25.50 & 21.26 & 18.11 & 11.21 & 19.02 \\
& WB-BLSTM1-SF \cite{meng2017} & 2.92 &    2.95 &    2.92 &    2.88 &    2.92  &
18.16 & 13.86 & 8.87 & 13.63 & 13.63 &
39.76 & 31.17 & 26.55 & 15.37 & 28.21 \\
& WB-BLSTM2-SF \cite{meng2017} & 3.02 &    3.02 &    2.97 &    2.94 &    2.99 &
12.94 & 10.47 & 7.30& 9.29 & 10.00 &
27.24 & 20.27 & 16.11 & 11.11 & 18.68 \\
2CH & BLSTM-MRM  &   2.77 &    2.81 &    2.82 &    2.77 &    2.79 &
 \textbf{10.60} &    5.68 &    5.12 &    6.33 & 6.93   &
\textbf{18.46} &   10.52 &    \textbf{9.59} &    \textbf{7.71} &  \textbf{11.57}    \\
& BLSTM-CC  &   2.90  &  2.94 &    2.93 &   2.89 &     2.91 &
   11.70 &    6.10 &    5.26 &    6.51 & 7.39 &
   20.88 &   10.85 &   10.97 &    8.32 & 12.75 \\  
& BLSTM-SF   &   2.88 &    2.94 &    2.88 &    2.86 &    2.89  & 
   11.45 &    6.06 &    5.18 &   6.23 & 7.23   &
   20.66 &   10.54 &   10.18 &    7.97 & 12.33     \\
   & BLSTM-SSF   & 2.75 &    2.78 &    2.79 &   2.74 &    2.77 &    
      10.97 &    \textbf{5.46} &    4.86 &    \textbf{6.13} & \textbf{6.86}  &
      19.37 &   \textbf{10.36} &    9.68 &    8.07 &  11.87    \\ \hline
& BeamformIt  \cite{anguera2007}&   1.77 &    2.19 &    2.31 &  2.10 &   2.09   &
 9.01 & 6.30 & 4.41 & 6.99 & 6.68 &
19.61 & 11.84 & 11.64 & 10.52 & 13.40 \\
&  NN-GEV \cite{heymann2016}  & 2.36 &    2.56 &   2.61 &    2.52 &    2.51  &
\textbf{5.41} & \textbf{4.03} & \textbf{3.60} & \textbf{4.32} & \textbf{4.34} &
11.39 & 6.57 & \textbf{7.32} & 6.95 & \textbf{8.06} \\
& WB-CRNN-MRM \cite{chakrabarty2019}  &   2.65 &    2.82 &    2.75 &    2.74 &    2.74 &
10.55 & 6.67 & 5.72 & 7.52 & 7.61 &
15.67 & 12.92 & 17.21 & 9.39 & 13.80  \\
& WB-BLSTM1-SF \cite{meng2017} & 2.82 &    2.93 &    2.88 &    2.85 &    2.87  &
13.32 & 10.60 & 7.27 & 10.66 & 10.46 &
30.53 & 24.34 & 20.95 & 13.30 & 22.28 \\
& WB-BLSTM2-SF \cite{meng2017} & 2.93 &    2.99 &    2.92 &    2.93 &    2.94  &
10.28 & 7.57 & 6.33 & 8.21 & 8.10 &
21.91 & 17.58 & 15.92 & 10.57 & 16.49 \\
4CH & BLSTM-MRM & 2.81  &  2.88 &    2.82 &    2.81 &    2.83  &
7.41 &    4.07 &    4.24 &    4.59  & 5.08 &
\textbf{10.44} &    6.77 &   11.29 &    6.55 & 8.76   \\
& BLSTM-CC  &    2.91 &    2.96 &    2.89 &   2.87 &    2.91 &
    6.62 &    4.34 &    4.22 &    4.72 & 4.97 &
11.31 &    6.97 &   10.33 &    6.40 & 8.75  \\
& LSTM-SF     & 2.71 & 2.78 &  2.76 &    2.74 &  2.75 & 
7.26 &    4.62 &    4.39 &    5.32 & 5.40 &
11.92 &    7.48 &   11.11 &    6.80 & 9.32 \\
& BLSTM-SF   &    2.89 &  2.93  &  2.89 &  2.86 &  2.89 & 
6.50 &    4.29 &    4.16 &   4.65  & 4.90 &
10.93 &    6.80 &   10.22 &    6.26 &   8.55            \\ 
& BLSTM-SSF   &  2.82 &    2.87 &    2.82 &    2.81 &    2.83 & 
    6.40 &    \textbf{4.03} &    3.92 &   4.53  &4.72 &
11.23 &    \textbf{6.17} &    9.59 &    \textbf{5.81} & 8.20  
\end{tabular}
\end{table*}   

%It is inconsistent to the belief in the community that TF masking methods are not suitable for ASR, our experiments show that the proposed narrow-band masking methods are able to efficiently improve the ASR performance. Therefore, we think, instead of the nonlinear operation of masking, what actually harms the ASR performance is the structured signal artifacts presented in the full-band techniques.   

\section{Conclusions}
\label{sec4}

In this paper we proposed a narrow-band deep filtering method to address the problem of multichannel speech enhancement. Unsupervised methods, such as spectral subtraction or spatial filtering, have shown some advantages of narrow-band processing for discriminating between speech and noise. The proposed LSTM-based method is able to exploit rich narrow-band features and it outperforms the methods mentioned above. Interestingly, narrow-band LSTM preserves one of the most prominent merits of unsupervised models, namely it is agnostic to speaker identity and to noise type. 

Four targets were used for training: the magnitude ratio mask (MRM), the complex coefficients (CC), the spatial filter (SF) and the smoothed SF (SSF). Most of these targets had already been studied in the wide-band speech enhancement framework. However, estimating these targets in narrow-band has completely different theoretical bases and behaviours from the wide-band cases. We evaluated the proposed narrow-band deep filtering method in terms of both speech enhancement and speech recognition. CC and SF achieve better speech enhancement performance than MRM by recovering the complex spectra. This superiority is more prominent for the 4CH case, since which provides more spatial features that the estimation of complex spectra can rely on. As for ASR, compared to CC and SF, MRM performs better for the 2CH case and slightly worse for the 4CH case, which shows that MRM is a proper target since ASR normally uses the (log-)magnitude feature. SSF notably improves the ASR performance over SF by temporally smoothing the spatial filter, while at the cost of worse speech enhancement performance. Compared with the state-of-the-art methods, the proposed methods achieve much better speech enhancement performance than the beamformers and the wide-band methods with relative small networks, i.e. WB-CRNN-MRM \cite{chakrabarty2019} and WB-BLSTM1-SF \cite{meng2017}. The wide-band method takes as input/output the high-dimensional full-band spectra, and needs a much larger network to properly learn useful informations, e.g. WB-BLSTM2-SF \cite{meng2017} with 54.6 M parameters (for reference the proposed network has 1.2 M parameters) achieves better speech enhancement performance than the proposed method, but the performance gap is not very big for the 4CH case. The wide-band methods don't work well for ASR due to their wide-band processing artifacts. The proposed methods achieve lower WERs than other methods for the 2CH case. The proposed BLSTM-SSF network achieves comparable WERs with the advanced beamforming technique, i.e. NN-GEV \cite{heymann2016}, especially for the Eval data.

It is interesting to note that by ignoring wide-band patterns, the proposed model has several merits: there is a large reduction in both the number of network parameters and the amount of training dataset, it has excellent generalization capabilities, and it avoids wide-band processing artifacts. It is however true that wide-band patterns contain interesting features that are not used with narrow-band models and which are worth to be included in order to further improve the performance of the proposed model. Therefore, it would be interesting to investigate new architectures that can incorporate wide-band features while preserving the advantages of narrow-band models, most notably their excellent generalization capabilities and their robustness against wide-band processing artifacts.

%In this work, the reverberation effect is not taken into account. The training speech, i.e. BTH speech, is inconsistent to the REAL test speech in which reverberation presents, especially in the BUS environments. However, the network still preforms quite well. In this experiment, it is infeasible to evaluate how the network treats reverberation due to the lack of reference clean speech. The dereverberation topic will be studied in the future. 
\balance

% -------------------------------------------------------------------------
% Either list references using the bibliography style file IEEEtran.bst
\bibliographystyle{IEEEtran}
\end{document}